\input harvmac
\input amssym
\input epsf

\def\unit{\relax{\rm 1\kern-.26em I}}
\def\nada{\relax{\rm 0\kern-.30em l}}

%\draftmode

%\def\Omega{\rho,\sigma,\nu  }

\def\det{{\rm det}}

%% MACROS
\noblackbox
\def\IL{\relax{\rm I\kern-.18em L}}
\def\IH{\relax{\rm I\kern-.18em H}}
\def\IR{\relax{\rm I\kern-.18em R}}
\def\IC{\relax\hbox{$\inbar\kern-.3em{\rm C}$}}
\def\IZ{\relax\ifmmode\mathchoice
{\hbox{\cmss Z\kern-.4em Z}}{\hbox{\cmss Z\kern-.4em Z}}
{\lower.9pt\hbox{\cmsss Z\kern-.4em Z}} {\lower1.2pt\hbox{\cmsss
Z\kern-.4em Z}}\else{\cmss Z\kern-.4em Z}\fi}
\def\CM {{\cal M}}
\def\CN {{\cal N}}

\def\CO {{\cal O}}

\def\CA{{\cal A}}

%% MORE MACROS
\def\CM {{\cal M}}
\def\CN {{\cal N}}

\def\CO {{\cal O}}

\def\det{{\rm det}}

\font\manual=manfnt \def\dbend{\lower3.5pt\hbox{\manual\char127}}

\def\IZ{\relax\ifmmode\mathchoice
{\hbox{\cmss Z\kern-.4em Z}}{\hbox{\cmss Z\kern-.4em Z}}
{\lower.9pt\hbox{\cmsss Z\kern-.4em Z}} {\lower1.2pt\hbox{\cmsss
Z\kern-.4em Z}}\else{\cmss Z\kern-.4em Z}\fi}
\def\half {{1\over 2}}

\def\rt2{\sqrt{2}}
\def\irt2{{1\over\sqrt{2}}}

\def\hat{\widehat}
%  \slashchar puts a slash through a character to represent contraction
%  with Dirac matrices. Use \not instead for negation of relations, and use
%  \hbar for hbar.
\def\slashchar#1{\setbox0=\hbox{$#1$}           % set a box for #1
   \dimen0=\wd0                                 % and get its size
   \setbox1=\hbox{/} \dimen1=\wd1               % get size of /
   \ifdim\dimen0>\dimen1                        % #1 is bigger
      \rlap{\hbox to \dimen0{\hfil/\hfil}}      % so center / in box
      #1                                        % and print #1
   \else                                        % / is bigger
      \rlap{\hbox to \dimen1{\hfil$#1$\hfil}}   % so center #1
      /                                         % and print /
   \fi}

\def\foursqr#1#2{{\vcenter{\vbox{
    \hrule height.#2pt
    \hbox{\vrule width.#2pt height#1pt \kern#1pt
    \vrule width.#2pt}
    \hrule height.#2pt
    \hrule height.#2pt
    \hbox{\vrule width.#2pt height#1pt \kern#1pt
    \vrule width.#2pt}
    \hrule height.#2pt
        \hrule height.#2pt
    \hbox{\vrule width.#2pt height#1pt \kern#1pt
    \vrule width.#2pt}
    \hrule height.#2pt
        \hrule height.#2pt
    \hbox{\vrule width.#2pt height#1pt \kern#1pt
    \vrule width.#2pt}
    \hrule height.#2pt}}}}
\def\psqr#1#2{{\vcenter{\vbox{\hrule height.#2pt
    \hbox{\vrule width.#2pt height#1pt \kern#1pt
    \vrule width.#2pt}
    \hrule height.#2pt \hrule height.#2pt
    \hbox{\vrule width.#2pt height#1pt \kern#1pt
    \vrule width.#2pt}
    \hrule height.#2pt}}}}
\def\sqr#1#2{{\vcenter{\vbox{\hrule height.#2pt
    \hbox{\vrule width.#2pt height#1pt \kern#1pt
    \vrule width.#2pt}
    \hrule height.#2pt}}}}
\def\square{\mathchoice\sqr65\sqr65\sqr{2.1}3\sqr{1.5}3}

\def\figin{\epsfcheck\figin}\def\figins{\epsfcheck\figins}
\def\epsfcheck{\ifx\epsfbox\UnDeFiNeD
\message{(NO epsf.tex, FIGURES WILL BE IGNORED)}
\gdef\figin##1{\vskip2in}\gdef\figins##1{\hskip.5in}% blank space instead
\else\message{(FIGURES WILL BE INCLUDED)}%
\gdef\figin##1{##1}\gdef\figins##1{##1}\fi}
\def\DefWarn#1{}
\def\figinsert{\goodbreak\midinsert}
\def\ifig#1#2#3{\DefWarn#1\xdef#1{fig.~\the\figno}
\writedef{#1\leftbracket fig.\noexpand~\the\figno}%
\figinsert\figin{\centerline{#3}}\medskip\centerline{\vbox{\baselineskip12pt
\advance\hsize by -1truein\noindent\footnotefont{\bf
Fig.~\the\figno:\ } \it#2}}
\bigskip\endinsert\global\advance\figno by1}

%\IntriligatorJJ
\lref\IntriligatorJJ{
  K.~A.~Intriligator, B.~Wecht,
  ``The Exact Superconformal R Symmetry Maximizes a,''
Nucl.\ Phys.\  {\bf B667}, 183-200 (2003). [hep-th/0304128].
%%CITATION = hep-th/0304128%%
}

%\SeibergPQ
\lref\SeibergPQ{
  N.~Seiberg,
  ``Electric - Magnetic Duality in Supersymmetric nonAbelian Gauge Theories,''
Nucl.\ Phys.\  {\bf B435}, 129-146 (1995). [hep-th/9411149].
%%CITATION = hep-th/9411149%%
}

%\KutasovIY
\lref\KutasovIY{
  D.~Kutasov, A.~Parnachev, D.~A.~Sahakyan,
 ``Central charges and U(1)(R) symmetries in N=1 superYang-Mills,''
JHEP {\bf 0311}, 013 (2003). [hep-th/0308071].
%%CITATION = hep-th/0308071%%
}

%\IntriligatorMI
\lref\IntriligatorMI{
  K.~A.~Intriligator, B.~Wecht,
  ``RG fixed points and flows in SQCD with adjoints,''
Nucl.\ Phys.\  {\bf B677}, 223-272 (2004). [hep-th/0309201].
%%CITATION = hep-th/0309201%%
}

%\WittenTW
\lref\WittenTW{
  E.~Witten,
  ``Global Aspects of Current Algebra,''
Nucl.\ Phys.\  {\bf B223}, 422-432 (1983).
%%CITATION = PRINT-83-0262 (PRINCETON)%%
}

%\FrishmanDQ
\lref\FrishmanDQ{
  Y.~Frishman, A.~Schwimmer, T.~Banks, S.~Yankielowicz,
  ``The Axial Anomaly and the Bound State Spectrum in Confining Theories,''
Nucl.\ Phys.\  {\bf B177}, 157 (1981).
%%CITATION = WIS-80/27-Ph%%
}

%\AppelquistHR
\lref\AppelquistHR{
  T.~Appelquist, A.~G.~Cohen, M.~Schmaltz,
  ``A New Constraint on Strongly Coupled Gauge Theories,''
Phys.\ Rev.\  {\bf D60}, 045003 (1999). [arXiv:hep-th/9901109
[hep-th]].
%%CITATION = SLAC-PUB-8045%%
}
%\AnselmiAM
\lref\AnselmiAM{
  D.~Anselmi, D.~Z.~Freedman, M.~T.~Grisaru, A.~A.~Johansen,
  ``Nonperturbative Formulas for Central Functions of Supersymmetric Gauge Theories,''
Nucl.\ Phys.\  {\bf B526}, 543-571 (1998). [hep-th/9708042].
%%CITATION = hep-th/9708042%%
}

%\AnselmiYS
\lref\AnselmiYS{
  D.~Anselmi, J.~Erlich, D.~Z.~Freedman, A.~A.~Johansen,
  ``Positivity Constraints on Anomalies in Supersymmetric Gauge Theories,''
Phys.\ Rev.\  {\bf D57}, 7570-7588 (1998). [hep-th/9711035].
%%CITATION = hep-th/9711035%%
}

%\CardyCWA
\lref\CardyCWA{
  J.~L.~Cardy,
  ``Is There a c Theorem in Four-Dimensions?,''
Phys.\ Lett.\  {\bf B215}, 749-752 (1988). }

%\OsbornTD
\lref\OsbornTD{
  H.~Osborn,
  ``Derivation of a Four-Dimensional c Theorem,''
Phys.\ Lett.\  {\bf B222}, 97 (1989).
%%CITATION = DAMTP-89-3%%
}

%\KutasovIY
\lref\KutasovIY{
  D.~Kutasov, A.~Parnachev, D.~A.~Sahakyan,
  ``Central Charges and U(1)(R) Symmetries in N=1 SuperYang-Mills,''
JHEP {\bf 0311}, 013 (2003). [hep-th/0308071].
%%CITATION = hep-th/0308071%%
}

%\IntriligatorMI
\lref\IntriligatorMI{
  K.~A.~Intriligator, B.~Wecht,
  ``RG Fixed Points and Flows in SQCD with Adjoints,''
Nucl.\ Phys.\  {\bf B677}, 223-272 (2004). [hep-th/0309201].
%%CITATION = hep-th/0309201%%
}

%\JackEB
\lref\JackEB{
  I.~Jack, H.~Osborn,
  ``Analogs for the C Theorem for Four-dimensional Renormalizable Field Theories,''
Nucl.\ Phys.\  {\bf B343}, 647-688 (1990).
%%CITATION = DAMTP-90-02%%
}

%\ZamolodchikovGT
\lref\ZamolodchikovGT{
  A.~B.~Zamolodchikov,
  ``Irreversibility of the Flux of the Renormalization Group in a 2D Field Theory,''
JETP Lett.\  {\bf 43}, 730-732 (1986). }

%\WessYU
\lref\WessYU{
  J.~Wess, B.~Zumino,
  ``Consequences of Anomalous Ward Identities,''
Phys.\ Lett.\  {\bf B37}, 95 (1971). }

%\SchwimmerZA
\lref\SchwimmerZA{
  A.~Schwimmer, S.~Theisen,
  ``Spontaneous Breaking of Conformal Invariance and Trace Anomaly Matching,''
Nucl.\ Phys.\  {\bf B847}, 590-611 (2011). [arXiv:1011.0696
[hep-th]].
%%CITATION = arXiv:1011.0696%%
}

%\PhamCR
\lref\PhamCR{
  T.~N.~Pham, T.~N.~Truong,
  ``Evaluation of the Derivative Quartic Terms of The Meson Chiral Lagrangian from Forward Dispersion Relation,''
Phys.\ Rev.\  {\bf D31}, 3027 (1985).
%%CITATION = Print-85-0588 (ECOLE POLY)%%
}

%\AdamsSV
\lref\AdamsSV{
  A.~Adams, N.~Arkani-Hamed, S.~Dubovsky, A.~Nicolis, R.~Rattazzi,
  ``Causality, Analyticity and an IR Obstruction to UV Completion,''
JHEP {\bf 0610}, 014 (2006). [hep-th/0602178].
%%CITATION = hep-th/0602178%%
}

%\DineSW
\lref\DineSW{
  M.~Dine, G.~Festuccia, Z.~Komargodski,
  ``A Bound on the Superpotential,''
JHEP {\bf 1003}, 011 (2010). [arXiv:0910.2527 [hep-th]].
%%CITATION = arXiv:0910.2527%%
}

%\CappelliYC
\lref\CappelliYC{
  A.~Cappelli, D.~Friedan, J.~I.~Latorre,
  ``C Theorem and Spectral Representation,''
Nucl.\ Phys.\  {\bf B352}, 616-670 (1991).
%%CITATION = RU-90-43%%
}

%\FrishmanDQ
\lref\FrishmanDQ{
  Y.~Frishman, A.~Schwimmer, T.~Banks, S.~Yankielowicz,
  ``The Axial Anomaly and the Bound State Spectrum in Confining Theories,''
Nucl.\ Phys.\  {\bf B177}, 157 (1981).
%%CITATION = WIS-80/27-Ph%%
}

%\WittenTW
\lref\WittenTW{
  E.~Witten,
  ``Global Aspects of Current Algebra,''
Nucl.\ Phys.\  {\bf B223}, 422-432 (1983).
%%CITATION = PRINT-83-0262 (PRINCETON)%%
}

%\BirrellIX
\lref\BirrellIX{
  N.~D.~Birrell, P.~C.~W.~Davies,
  ``Quantum Fields in Curved Space,''
Cambridge, Uk: Univ. Pr. ( 1982) 340p.
}

%\MyersTJ
\lref\MyersTJ{
  R.~C.~Myers, A.~Sinha,
 ``Holographic c-Theorems in Arbitrary Dimensions,''
JHEP {\bf 1101}, 125 (2011).
[arXiv:1011.5819 [hep-th]].
%%CITATION = arXiv:1011.5819%%
}

%\DuffWM
\lref\DuffWM{
  M.~J.~Duff,
  ``Twenty Years of the Weyl Anomaly,''
Class.\ Quant.\ Grav.\  {\bf 11}, 1387-1404 (1994).
[hep-th/9308075].
%%CITATION = hep-th/9308075%%
}

%\BuchbinderJN
\lref\BuchbinderJN{
  I.~L.~Buchbinder, S.~M.~Kuzenko, A.~A.~Tseytlin,
 ``On Low-Energy Effective Actions in N=2, N=4 Superconformal Theories in Four-Dimensions,''
Phys.\ Rev.\  {\bf D62}, 045001 (2000).
[hep-th/9911221].
%%CITATION = hep-th/9911221%%
}

%\JafferisZI
\lref\JafferisZI{
  D.~L.~Jafferis, I.~R.~Klebanov, S.~S.~Pufu, B.~R.~Safdi,
  ``Towards the F-Theorem: N=2 Field Theories on the Three-Sphere,''
JHEP {\bf 1106}, 102 (2011).
[arXiv:1103.1181 [hep-th]].
%%CITATION = arXiv:1103.1181%%
}

%\RiegertKT
\lref\RiegertKT{
  R.~J.~Riegert,
 ``A Nonlocal Action for the Trace Anomaly,''
Phys.\ Lett.\  {\bf B134}, 56-60 (1984).
}

%\FradkinTG
\lref\FradkinTG{
  E.~S.~Fradkin, A.~A.~Tseytlin,
 ``Conformal Anomaly in Weyl Theory and Anomaly Free Superconformal Theories,''
Phys.\ Lett.\  {\bf B134}, 187 (1984).
%%CITATION = LEBEDEV-83-180%%
}

%\TS
\lref\TS{
  A.~Schwimmer, S.~Theisen,
 in preparation.}

%\KulaxiziJT
\lref\KulaxiziJT{
  M.~Kulaxizi, A.~Parnachev,
  ``Energy Flux Positivity and Unitarity in CFTs,''
Phys.\ Rev.\ Lett.\  {\bf 106}, 011601 (2011).
[arXiv:1007.0553 [hep-th]].
%%CITATION = arXiv:1007.0553%%
}

%\HofmanAR
\lref\HofmanAR{
  D.~M.~Hofman, J.~Maldacena,
  ``Conformal Collider Physics: Energy and Charge Correlations,''
JHEP {\bf 0805}, 012 (2008).
[arXiv:0803.1467 [hep-th]].
%%CITATION = arXiv:0803.1467%%
}

%\FradkinYW
\lref\FradkinYW{
  E.~S.~Fradkin, G.~A.~Vilkovisky,
  ``Conformal Off Mass Shell Extension and Elimination of Conformal Anomalies in Quantum Gravity,''
Phys.\ Lett.\  {\bf B73}, 209-213 (1978).
}

%\DeserYX
\lref\DeserYX{
  S.~Deser, A.~Schwimmer,
  ``Geometric Classification of Conformal Anomalies in Arbitrary Dimensions,''
Phys.\ Lett.\  {\bf B309}, 279-284 (1993).
[hep-th/9302047].
%%CITATION = hep-th/9302047%%
}

%\BonoraCQ
\lref\BonoraCQ{
  L.~Bonora, P.~Pasti, M.~Bregola,
  ``Weyl Cocycles,''
Class.\ Quant.\ Grav.\  {\bf 3}, 635 (1986).
%%CITATION = DFPD-28/85%%
}

%\NirSV
\lref\NirSV{
  Y.~Nir,
 ``Infrared Treatment of Higher Anomalies and Their Consequences,''
Phys.\ Rev.\  {\bf D34}, 1164-1168 (1986).
%%CITATION = PRINT-87-0059%%
}

%\DistlerIF
\lref\DistlerIF{
  J.~Distler, B.~Grinstein, R.~A.~Porto, I.~Z.~Rothstein,
  ``Falsifying Models of New Physics via WW Scattering,''
Phys.\ Rev.\ Lett.\  {\bf 98}, 041601 (2007). [hep-ph/0604255].
%%CITATION = hep-ph/0604255%%
}

%\FreedmanGP
\lref\FreedmanGP{
  D.~Z.~Freedman, S.~S.~Gubser, K.~Pilch, N.~P.~Warner,
  ``Renormalization group flows from holography supersymmetry and a c theorem,''
Adv.\ Theor.\ Math.\ Phys.\  {\bf 3}, 363-417 (1999).
[hep-th/9904017].
%%CITATION = hep-th/9904017%%
}

%\GirardelloBD
\lref\GirardelloBD{
  L.~Girardello, M.~Petrini, M.~Porrati, A.~Zaffaroni,
  ``The Supergravity dual of N=1 superYang-Mills theory,''
Nucl.\ Phys.\  {\bf B569}, 451-469 (2000). [hep-th/9909047].
%%CITATION = hep-th/9909047%%
}

%%%%%%%%%%%%%%%%%%%%%%%%%%%%%%%%%%%%%%%%%%%%%%%%%%%%%%%%%%%%%%%%%%%%%%%%%%%%%%%%%%%%%%%%%%%%%%%%%%%%\rightline{CERN-PH-TH/2011-112} \Title{
\rightline{WIS/06/11-AUG-DPPA} \Title{
%\rightline{hep-th/yymmnnn}
} {\vbox{\centerline{ On Renormalization Group Flows in Four
Dimensions } }}
\medskip

\centerline{\it Zohar Komargodski $^{\spadesuit\heartsuit}$ and
Adam Schwimmer $^{\spadesuit}$}
\bigskip
\centerline{$^\spadesuit$ Weizmann Institute of Science, Rehovot
76100, Israel}
 \centerline{$^\heartsuit$
Institute for Advanced Study, Princeton, NJ 08540, USA}

\smallskip

\vglue .3cm
\bigskip
\bigskip
\bigskip
\noindent We discuss some general aspects of renormalization group
flows in four dimensions. Every such flow can be reinterpreted in
terms of a spontaneously broken conformal symmetry. We analyze in
detail the consequences of trace anomalies for the effective
action of the Nambu-Goldstone boson of broken conformal symmetry.
While the $c$-anomaly is algebraically trivial, the $a$-anomaly is
``non-Abelian,'' and leads to a positive-definite universal
contribution to the $S$-matrix element of $2\rightarrow2$ dilaton
scattering. Unitarity of the $S$-matrix results in a monotonically
decreasing function that interpolates between the Euler anomalies
in the ultraviolet and the infrared, thereby establishing the
$a$-theorem.

\Date{Aug 2011}
%\draftmode

\newsec{Introduction}

One of the fundamental questions about quantum field theory is
whether the Renormalization Group (RG) flux is reversible. Namely,
if there exist two conformal field theories $A,B$ such that
one can flow from $A$ to $B$ but also from $B$ to $A$. The answer
in two-dimensional field theories has been long known to be
negative~\ZamolodchikovGT, but a corresponding result for
four-dimensional field theories has so far eluded us.

In the case of two space-time dimensions, Zamolodchikov has
established the existence of a monotonically decreasing function,
$C$, interpolating between the central charges of the UV and IR
CFTs. This not only proves that the RG flux is irreversible, but
also provides an effective measure for the number of degrees of
freedom, such that as we integrate out high momentum modes this
number decreases.

There are several conceivable ways to generalize
to 4d. One such proposal by Cardy~\CardyCWA\ (see
also~\refs{\OsbornTD,\JackEB}) states that one should consider the
integral of the energy-momentum tensor over the four-sphere
\eqn\cardprop{a\sim \int_{S^4} \langle T_\mu^\mu\rangle~.}
Cardy's conjecture is that the quantity~\cardprop\ decreases as we flow.\foot{Another proposal for a measure of degrees of freedom in four
dimensions is due to Appelquist-Cohen-Schmaltz~\AppelquistHR.
We will have nothing to add to this conjecture  here.}

The trace of the stress tensor in four-dimensional conformal field
theories is nonzero in curved spaces (see~\DuffWM\ and references
therein) due to the trace anomalies $a$ and $c$
\eqn\traceanomalies{T_\mu^\mu=aE_4-cW_{\mu\nu\rho\sigma}^2~,}
where $E_4$ is the Euler density and $W_{\mu\nu\rho\sigma}^2$ is
the Weyl tensor squared.\foot{Interesting bounds on the values $a$
and $c$ may assume at fixed points have been devised
recently~\refs{\HofmanAR,\KulaxiziJT}.} Hence, in the UV and IR
CFTs the integral over the four-sphere~\cardprop\  isolates the
$a$-anomalies $a_{UV}$ and $a_{IR}$, respectively. Cardy's
conjecture thus implies that, in particular,
\eqn\athm{a_{IR}<a_{UV}~.}

Generally speaking, one can study the quantities $a_{UV}$,
$a_{IR}$ only in a limited set of examples. Perturbative fixed
points serve as one realm of theories where~\athm\ can be put to
test, and it has always been found to hold. With supersymmetry the
situation becomes much better and many more examples can be
examined. This is due to duality~\SeibergPQ, the important
relation between $a$ and the superconformal
$R$-symmetry~\refs{\AnselmiAM,\AnselmiYS}, and the breakthrough of
a-maximization~\IntriligatorJJ\ (see also
\refs{\KutasovIY,\IntriligatorMI}). All the known examples have
yielded results consistent with~\athm. The $a$-theorem has been
also widely discussed in the context of holography, starting
from~\refs{\FreedmanGP,\GirardelloBD}. See also the recent
illuminating study in~\MyersTJ.

Needless to say, it is important to know if~\athm\ is indeed a
true property of quantum field theory in four dimensions. Very
much like 't Hooft's anomaly matching, this can lead to strong
constraints on the dynamics of gauge theories, the patterns of
symmetry breaking, and other questions that can be relevant to
particle physics in the near future.

For chiral symmetries that are conserved along the flow, 't Hooft argued that all the anomalies should match.
Recall that one cancels the anomalies by introducing
very weakly interacting (spectator) fields into the system and
then argues that consistency of the IR theory implies that the
anomalies in the IR should reproduce those in the UV.

One cannot quite repeat this argument for the conformal (alternatively, Weyl)
symmetry, since every RG flow violates it explicitly. In the deep UV and the deep IR the symmetries of the problem are enhanced to the conformal group,  but the conformal symmetry is broken throughout the bulk of the flow. Therefore, the anomalies at high energies generally differ from those at low energies. In this sense, that there is any possible regularity in these anomalies~\athm\ is surprising.

In this paper we develop a formalism which allows to follow this violation of conformal symmetry along the flow. As a result, we prove the inequality~\athm\ for all  unitary RG flows. We also establish a monotonically decreasing interpolating function between
$a_{UV}$ and $a_{IR}$.

The main idea is to use a dilaton spectator field in order to
reinterpret every massive RG flow as if it results from
spontaneously broken conformal symmetry. This can be done while
keeping the dilaton fluctuations arbitrarily weakly coupled to the
matter theory. We then study carefully the effective action of the
dilaton field and show that one particular four-derivative term in
this theory is related to the $a$-anomaly. This special term in
the dilaton effective action is reminiscent of the topological
term in pion physics~\refs{\WessYU,\WittenTW}. Finally, we show
that this term can be isolated by computing a $2\rightarrow2$
$S$-matrix element and it satisfies a positive-definite dispersion
relation, establishing our main claim. This also gives rise to a
monotonically decreasing function interpolating between $a_{UV}$
and $a_{IR}$. Morally speaking, our dilaton field can be thought
of as a cousin of 't Hooft's spectators, however, many other
aspects of the analysis have no analogs in the argument for
anomaly matching of global symmetries.

The plan of this paper is as follows. In section~2 we discuss the rules of writing diff$\times$Weyl invariant actions for the dilaton field in general curved backgrounds. We also discuss the anomaly functional and show that there is a particular four-derivative term which is uniquely determined by the $a$-anomaly. (More precisely, $a$ fixes the result of a particular low energy $S$-matrix calculation.)
As a warm-up exercise that highlights some of the important ingredients in our construction, we prove a special case of the $a$-theorem in section~3. In section~4 we explain how every massive RG flow can be reinterpreted as coming from spontaneously broken conformal symmetry and prove the strongest version of the $a$-theorem. Various open questions and further research directions are briefly discussed in section~5. Some of the conventions used throughout this paper are summarized in appendix~A. In appendix~B we illustrate the methods of section~4 for the flow of a free massive field.

\newsec{The Theory of the Dilaton: Invariant Terms and Anomalous Functionals}

\subsec{Invariant Terms} Consider a spontaneously broken CFT. Then, by the Nambu-Goldstone theorem, there is a massless particle, the dilaton $\tau$. Effective
actions for it follow the same rules as in any theory of
spontaneously broken symmetry. (One can also use current algebra techniques to derive the same results.)
It is easy to organize such actions
by introducing a space-time metric $g_{\mu\nu}$ and demanding that
the theory is invariant under diff$\times$Weyl transformations,
where Weyl transformations act as
\eqn\weyltran{g_{\mu\nu}\longrightarrow
e^{2\sigma}g_{\mu\nu}~,\qquad \tau\longrightarrow \tau+\sigma~.}
Diffeomorphisms act as usual, with the dilaton being a space-time
scalar. We will often denote $\hat g=e^{-2\tau}g_{\mu\nu}$. The
combination $\hat g$ transforms as a metric under diffeomorphisms and is Weyl invariant.

The most general theory up to two derivatives is:
\eqn\effact{f^2\int d^4x\sqrt{-\det \hat g}\left(\Lambda+{1\over
6}\hat R\right)~,} where we have defined $\hat R=\hat
g^{\mu\nu}R_{\mu\nu}[\hat g]$. $f$ is the ``decay constant'' of the
spontaneously broken conformal theory.
The cosmological constant term $\Lambda$ leads to a
scale-invariant potential for the dilaton. There is nothing wrong
with it by itself, except that if the dilaton really comes from
spontaneously broken conformal symmetry the vacuum degeneracy
cannot be lifted and hence $\Lambda=0$. (This is the well-known statement that the cosmological constant is zero in vacua that break the conformal symmetry spontaneously.)

Since we are ultimately interested in the Minkowskian theory, let
us evaluate the kinetic term with $g_{\mu\nu}=\eta_{\mu\nu}$.
Using integration by parts we get \eqn\kineticflat{S=f^2\int d^4x
e^{-2\tau}(\del \tau)^2 ~.} This describes a free massless
particle, albeit in a somewhat strange choice of variables.
One can use the field redefinition
\eqn\canonical{\varphi=1-e^{-\tau}} to bring the kinetic term into canonical form.
The utility of the variable $\tau$ will be seen later. The equation of
motion is \eqn\eom{\square \ \tau = (\del \tau)^2~.}

One can also study terms in the effective action with more
derivatives. With four derivatives, one has three independent
(dimensionless) coefficients \eqn\allfour{\int d^4x\sqrt{-\hat
g}\left(\kappa_1\hat R^2+\kappa_2\hat R_{\mu\nu}^2+\kappa_3\hat
R_{\mu\nu\rho\sigma}^2\right)~.} It is implicit that indices are
raised and lowered with $\hat g$. (We have not included the Pontryagin term as well as $\hat \square \ \hat R$ in~\allfour\ since they both integrate to zero.) This basis of interactions is
somewhat inconvenient for our purposes. Recall the expressions for
the Euler density $\sqrt{-g} E_4$ and the Weyl tensor squared
\eqn\recall{E_4=R_{\mu\nu\rho\sigma}^2-4 R_{\mu\nu}^2+R^2~,\qquad
W_{\mu\nu\rho\sigma}^2=R_{\mu\nu\rho\sigma}^2-2 R_{\mu\nu}^2+{1\over 3}R^2
~.} We can thus choose instead of~\allfour\
a different parameterization \eqn\allfouri{\int d^4x\sqrt{-\hat
g}\left(\kappa'_1\hat R^2+\kappa'_2\hat E_4+\kappa'_3\hat
W_{\mu\nu\rho\sigma}^2\right)~.} We immediately see that the $\kappa'_2$ term
is a total derivative. If we set $g_{\mu\nu}=\eta_{\mu\nu}$,
then $\hat g_{\mu\nu}=e^{-2\tau}\eta_{\mu\nu}$ is conformal to the
flat metric and hence also the $\kappa'_3$ term does not play any role as
far as the dilaton interactions in flat space are concerned. Consequently, terms in the flat space limit arise solely from $\hat R^2$. A
straightforward calculation yields  \eqn\kappaone{\int
d^4x\sqrt{-\hat g}\hat R^2\biggr|_{g_{\mu\nu}=\eta_{\mu\nu}}=36\int d^4x\left(\square \ \tau-(\del
\tau)^2\right)^2~. }

The combination on the right hand side of~\kappaone\  vanishes by
the equation of motion of the two-derivative theory~\eom.
(Equivalently, using the variable $\varphi$ defined in~\canonical\
the left hand side of~\kappaone\ is seen to be proportional to
$(\square  \ \varphi)^2$, which vanishes on-shell.) The mistake in
using the zeroth order equation of motion is higher order in the
derivative expansion. Thus, the diff$\times$Weyl invariant terms
in the Lagrangian do not yield a genuine tree-level
four-derivative interaction. Or, said more invariantly, there is
no $s^2+t^2+u^2$ term in the low momentum expansion of the
scattering amplitude of four dilatons.\foot{A more conceptual way
of seeing that~\kappaone\ had to vanish on-shell comes by noting
that, in the absence of the cosmological constant term in~\effact,
the equation of motion for the dilaton is just the trace of the
Einstein equation, hence, the Ricci scalar vanishes on-shell.}

\subsec{Anomalous Functionals}

We should also contemplate functionals of $g_{\mu \nu},\tau$
which are not diff$\times$Weyl invariant. This is important
because the physical theories we will discuss in the next sections
have trace anomalies and we will need to match these anomalies
with functionals of $g_{\mu\nu},\tau$.

The most general anomalous variation one needs to consider takes
the form \eqn\anomafun{\delta_\sigma S_{anomaly}=\int d^4
x\sqrt{-g}\sigma\left(c W_{\mu\nu\rho\sigma}^2- a E_4 + b'\square
R\right)~.} The question is then how to write a functional
$S_{anomaly}$ that reproduces this anomaly. (Note that
$S_{anomaly}$ is only defined modulo diff$\times$Weyl invariant
terms.) Without the field $\tau$ one must resort to non-local
expressions, but in the presence of the dilaton one has a local
action.

The term $b'$ is uninteresting to us because it can be accounted for
by a local functional that only depends on $g$ and not on $\tau$
(hence, $b'$ is not associated to an anomaly; it is simply a contact term in the two-point function of stress
tensors that we may or may not want to include). To verify this one
can easily check that \eqn\var{\delta_\sigma\int
R^2\sim \int \sigma \square R~.} Hence, the coefficient
$b'$ does not affect the matrix elements of $\tau$ in flat space.
We will henceforth set $b'=0$ in this paper. To reintroduce
$b'$ one simply adds $\sim \int d^4 x\sqrt {-g} R^2$  to the anomaly
functional.

It is a little tedious to solve~\anomafun, but the procedure is straightforward in principle. (See also~\refs{\FradkinYW\RiegertKT\FradkinTG-\SchwimmerZA} and references therein for other approaches to the problem.) We first replace
$\sigma$ on the right-hand side of~\anomafun\ with $\tau$
\eqn\firstguess{S_{anomaly}=\int d^4x \sqrt{- g}\tau\left( c
W_{\mu\nu\rho\sigma}^2-a E_4\right)+\cdots~.} While the variation of this includes the sought-after terms~\anomafun,  as the
$\cdots$ suggest, this cannot be the whole answer because the
object in parenthesis is not Weyl invariant. Hence, we need to
keep fixing this expression with more factors of $\tau$ until the
procedure terminates. Note that $\sqrt {-g}W_{\mu\nu\rho\sigma}^2$, being
the square of the Weyl tensor, is Weyl invariant, and hence we do
not need to add any fixes proportional to the $c$-anomaly
in~\firstguess. This makes the $c$-anomaly ``Abelian'' in some sense.
The ``non-Abelian'' structure coming from the Weyl variation of $E_4$
is the key to our construction. The $a$-anomaly is therefore quite
distinct algebraically from the $c$-anomaly.

The final expression for $S_{anomaly}$ is
\eqn\minimalanomaly{\eqalign{S_{anomaly}=&-a\int d^4x\sqrt{- g}\biggl(
 \tau E_4+ 4\bigl(R^{\mu\nu}-\half g^{\mu\nu}
R\bigr)\del_\mu \tau\del_\nu \tau-4(\del
\tau)^2\square\
 \tau+2(\del \tau)^4\biggr)\cr&+c\int d^4x \sqrt{-g}  \tau
W_{\mu\nu\rho\sigma}^2 ~.}}
Note that even when the metric is flat, self-interactions of the
dilaton survive. This is analogous to what happens in pion physics when the background gauge fields are set to zero~\refs{\WessYU,\WittenTW}. These interaction terms in flat space-time are forced on us by the ``non-Abelian'' structure of the
$a$-anomaly.

Since the self interactions
which survive $g_{\mu\nu}=\eta_{\mu\nu}$ are four-derivative
interactions, we can use the equation of motion~\eom\ to simplify
things.  Hence, we get that the
anomaly contribution is
\eqn\minimalanomalyi{\eqalign{S_{anomaly}\biggl|_{
g_{\mu\nu}=\eta_{\mu\nu}}\approx 2a\int d^4x (\del \tau)^4~.}}
Above, the~$\approx$ symbol means that we have used the leading order equation of motion. Physical observables such as the $S$-matrix are invariant under using the equations of motion, so we loose nothing by utilizing~\eom\ at the order of four derivatives.

Let us conclude what we have found in this and the previous
subsection. No diff$\times$Weyl invariant terms give rise to the
scattering of four dilatons at the order of four
derivatives. Such a contribution, however, arises from the anomaly
functional, and its coefficient is fixed by the $a$-anomaly.

This important conclusion is independent of whether or not there
is a $b'$ term in the anomaly~\anomafun, because $b'$ can always
be canceled by a local counterterm~\var\ that does not affect
dilaton scattering in flat space. It is also independent of our
choice of the anomaly functional~\minimalanomaly\ because all such
choices differ by diff$\times$Weyl invariant terms, and the latter
have no effect on flat-space dilaton scattering at the level of
four derivatives. We therefore arrived at a low energy theorem for dilaton scattering. (Analogous to some soft pion theorems.)

\newsec{The $a$-Theorem for Motion on the Moduli Space}
Consider a conformal field theory, CFT$_{UV}$ (with anomalies
$a_{UV}, c_{UV}$), which has a moduli space of vacua $\CM$. In all
of them but the one at the ``origin'' conformal symmetry is
spontaneously broken.  Let us study the physics in one of these
conformal symmetry-breaking vacua. Generally, at the scale of
breaking $f$ there are massive particles, but there can be some
massless particles too. In fact, the dilaton, which is the
Nambu-Goldstone boson of conformal symmetry breaking, must be
massless. In addition to it, there may be a nontrivial IR CFT,
denoted CFT$_{IR}$ (with anomalies $a_{IR}, c_{IR}$). The
situation is summarized in Fig.1.

\medskip
\epsfxsize=3.2in \centerline{\epsfbox{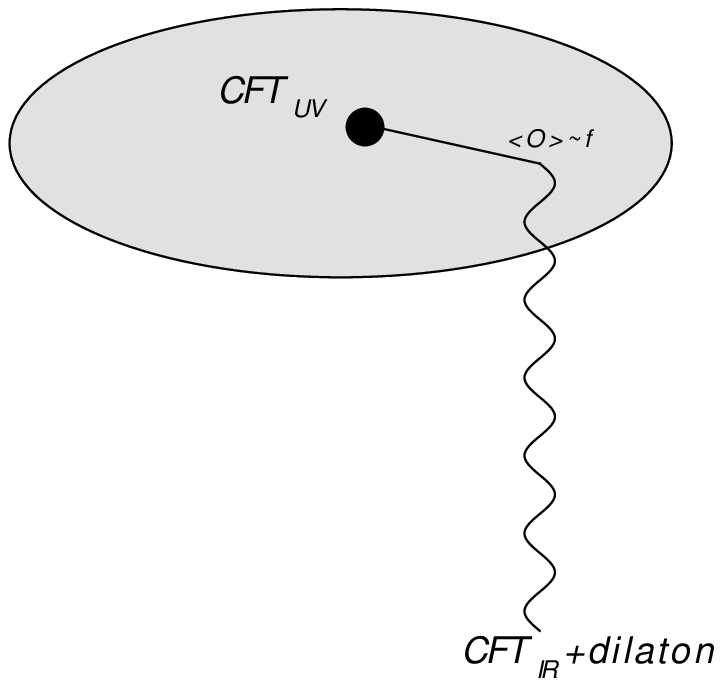}}
\noindent
Fig.1: The shaded region represents the moduli space of the UV CFT. An operator ${\cal O}$ obtains a VEV of order $f$ and breaks the conformal symmetry
spontaneously. This results in a flow to a low energy theory containing the massless dilaton and possibly a non-trivial conformal field theory, CFT$_{IR}$. \medskip

Moduli spaces of vacua might not be easy to find in non-supersymmetric theories, but such moduli spaces are ubiquitous in supersymmetric theories and the RG flows one can trigger by turning on VEVs for moduli may lead to intricate IR CFTs. It is therefore not at all obvious a priori that $a_{IR}<a_{UV}$ is satisfied.

In conformal field theory, whether it is spontaneously broken or not, the total energy-momentum tensor is traceless, in other words, it satisfies the operator equation $T_\mu^\mu=0$. Anomalies show up in curved space (alternatively, as contact terms in special correlation functions).  In this situation, there is no {\it operatorial} violation of the Ward identity $T_\mu^\mu=0$, only a {\it $c$-number} violation due to anomalies. Therefore, the total anomalies in the UV and IR must agree because of the usual arguments for anomaly matching~\SchwimmerZA. This does not mean that the $a$- and $c$-anomalies of CFT$_{UV,IR}$ match, rather, that the difference must  be compensated for by the dilaton.\foot{This situation is somewhat reminiscent of the role of the Liouville field in non-critical string theory.}

The matching of the total anomaly forces the effective action in the IR to take the form
\eqn\IRaction{\eqalign{S_{IR}[g_{\mu\nu}]&={\rm {CFT}}_{IR}[g_{
 \mu\nu}]+{1\over 6}f^2\int d^4x\sqrt{- \hat
g}\hat R+{\kappa\over 36}\int d^4x\sqrt {-\hat g}\hat
R^2+\kappa'\int d^4x\sqrt {-\hat g}\hat W_{\mu\nu\rho\sigma}^2\cr&
-(a_{UV}-a'_{IR})\int d^4x\sqrt{- g}\biggl(
 \tau E_4+ 4\bigl(R^{\mu\nu}-\half g^{\mu\nu}
R\bigr)\del_\mu \tau\del_\nu \tau-4(\del
\tau)^2\square\
 \tau+2(\del \tau)^4\biggr)\cr&+(c_{UV}-c'_{IR})\int d^4x \sqrt{-g}  \tau
W_{\mu\nu\rho\sigma}^2+\cdots~.}} The cosmological constant term
from~\effact\ is zero ($\Lambda=0$) since we are working on the
moduli space and hence there cannot be a potential for the
dilaton. The constant $f$  is the physical decay constant of the
dilaton. In addition, we have defined $a'_{IR}=a_{IR}+a_{scalar}$,
$c'_{IR}=c_{IR}+c_{scalar}$. The reason is that in addition to the
explicit tree-level contribution contained in the effective
action, the dilaton contributes to the total anomalies like any
massless scalar via loop diagrams.\foot{We thank R.~Myers for a
discussion.}

The different terms in the effective action~\IRaction\ may be
generated by integrating out the massive modes.  The most
important feature in~\IRaction\ is, of course, that the anomaly
functional comes with prescribed coefficients to match the total
anomalies.\foot{In general we need to cancel the $b'$
anomaly~\anomafun\ too. Even if we choose it to vanish in
CFT$_{UV}$, it would generally be nonzero in CFT$_{IR}$. This must
be taken into account by adding a term proportional to $\sqrt
gR^2$ in~\IRaction. This contribution does not affect our
discussion so we omit it. }

We would like to examine the effective action~\IRaction\ for a
flat space-time metric  $g_{\mu\nu}=\eta_{\mu\nu}$. The result is
\eqn\flatIR{\eqalign{S_{IR}&={\rm {CFT}}_{IR}\cr&+\int
d^4x\left(f^2e^{-2\tau}(\del\tau)^2+\kappa\left( \square \
\tau-(\del\tau)^2\right)^2
+(a_{UV}-a'_{IR})\left(4(\del\tau)^2\square \
\tau-2(\del\tau)^4\right) \right)+\cdots~.}}

We see that the difference between the $a$-anomalies
$a_{UV}-a'_{IR}$ appears in front of some specific four-derivative
terms, and other terms with four derivatives  appear to be
multiplied by an unknown coefficient $\kappa$. As we have
explained in section 2,  these two types of contributions are
neatly disentangled when one considers the $S$-matrix for
$\tau\tau$ scattering. The leading contribution to the scattering
amplitude is fixed by the difference of the $a$-anomalies (recall
the usual relation $s+t+u=0$) \eqn\scatampl{\CA(s,t)=
{a_{UV}-a'_{IR}\over f^4}\left(s^2+t^2+u^2\right)+\cdots~.} Higher
order contributions, encompassed by the~$\cdots$, may be
model-dependent.

It has been known for a while that low energy contributions to
scattering elements can be sometimes evaluated by dispersion
relations. This follows from analyticity. For instance, certain
subleading operators in the chiral Lagrangian are known to have
positive coefficients~\PhamCR. (See also~\DistlerIF\ for a
discussion in the context of $WW$ scattering.) This positivity
constraint has been also shown to be closely related to the
absence of low energy superluminal modes~\AdamsSV. Many other
constraints of this kind exist and have applications for a wide
range of problems, see for instance~\DineSW\ for one such example.

After using the equation of motion in~\flatIR, the quartic term
one remains with is proportional to $(a_{UV}-a'_{IR})(\del
\tau)^4$. This operator is the simplest example for the
superluminal behavior discussed in~\AdamsSV. The absence of
superluminal modes immediately implies that $a'_{IR}\leq a_{UV}$
(and thus $a_{IR}<a_{UV}$). We will now study more closely the
analytic structure and establish the stronger inequality $a'_{IR}<
a_{UV}$, along with constructing a sum rule for the difference
$a_{UV}-a'_{IR}$. This also leads to a monotonically decreasing
function that interpolates between $a_{UV}$ and $a'_{IR}$.

Consider the scattering of four dilatons, with momenta
$p_1,p_2,p_3,p_4$ such that they are all on-shell $p_i^2=0$ and of
course $\sum_{i=1}^4 p_i=0$. We assume that we are in the forward
limit $t=0$ where \eqn\forward{p_1=-p_3~,\qquad p_2=-p_4~.} The
amplitude for this scattering process thus becomes~\scatampl\
\eqn\amplitude{\CA(s)= {2(a_{UV}-a'_{IR})\over
f^4}s^{2}+\CO(s^4)~.}

The last step is to consider the amplitude $\CA/s^3$ and write a
dispersion relation for it. There are branch cuts both at positive
and negative $s$. Negative $s$ cuts correspond to physical states
in the $u$ channel, and the symmetry $s\leftrightarrow u$ renders
these contributions identical to the ones for positive $s$. In
addition, $\CA/s^3$ has a pole at the origin which selects the
coefficient $a_{UV}-a'_{IR}$. Hence, by closing the contour we can
write a dispersion relation: \eqn\disrel{a_{UV}-a'_{IR}={f^4\over
\pi}\int_{s'>0} ds'{Im \CA(s')\over s'^3}~.} Here $Im \CA(s')$
denotes the imaginary part of the amplitude. This discontinuity is
positive definite because it satisfies $Im \CA(s)=s\sigma(s)$,
where $\sigma(s)$ is the total cross section for the scattering of
$\tau\tau$. We conclude that $a'_{IR}<a_{UV}$. This also provides
a natural function that decreases along the flow. Define a
scale-dependent ``$a$-anomaly'' \eqn\decrease{a(\mu)\equiv
a_{UV}-{f^4\over \pi}\int_{s'>\mu} ds'{\sigma(s')\over s'^2}~.}
This decreases monotonically as a function of $\mu$ and
interpolates between $a_{UV}$ at $\mu\rightarrow\infty$ and
$a'_{IR}$ at $\mu\rightarrow 0$. This proves the $a$-theorem for
this class of theories.\foot{We have been a little cavalier in
manipulating the dispersion relation above, not explaining why it
converges. If the dispersion relation had diverged this would have
meant that the difference between the anomalies needs a
subtraction (i.e. a counterterm). This is clearly not the case
since $a_{UV}-a_{IR}$ is a physical quantity. }

\newsec{Renormalization Group Flows as Spontaneously Broken Conformal Symmetry and the $a$-Theorem}

In the previous section we have discussed renormalization group flows which are triggered by turning on VEVs for operators that parameterize the moduli space of some conformal field theory.
However, the most interesting case to consider is a CFT$_{UV}$ which is deformed by some relevant operator(s) $M^{4-\Delta}\CO_\Delta$. (The operator can also be marginally relevant, like the gauge coupling in QCD.) This sets off a flow to some IR physics, which in the deep low energy limit is described by a possibly nontrivial  CFT$_{IR}$.  We summarize this in Fig.2.
\medskip
\epsfxsize=0.9in \centerline{\epsfbox{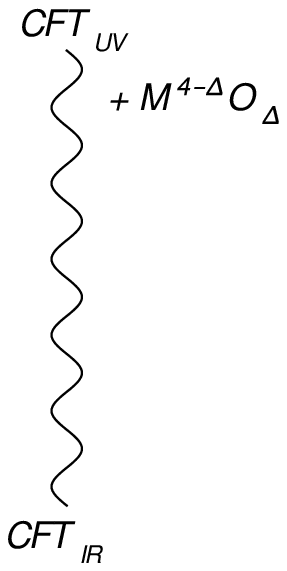}} \noindent
Fig.2: Starting from some CFT at high energies, we add a relevant
operator $\CO_\Delta$ and flow to a new CFT in the deep infrared.
\medskip

Consider the flow of the matter theory described in Fig.2. The matter theory is described by an action functional for some fields $\Phi_i$ and some mass parameters $M_i$
\eqn\mattertheory{S_{matter}=S_{matter}[\Phi_i,M_i]~.}
Upon coupling this theory to a background metric $g_{\mu\nu}$ the partition function is guaranteed to be diffeomorphism invariant by virtue of the conservation of the stress tensor. However, Weyl invariance is violated by the $a$- and $c$-anomalies~\traceanomalies, as well as the explicit mass parameters which induce a nonzero $(T^{matter})_\mu^\mu$ in flat space. The nonzero $(T^{(matter)})_\mu^\mu$ due to the mass parameters $M_i$ is referred to as  the {\it operatorial anomaly}, to be distinguished from the {\it $c$-number} anomalies which do not enter the operator equation for $(T^{matter})_\mu^\mu$ in flat space (the $c$-number anomalies only manifest themselves via contact terms in special correlation functions).

What precludes one from matching straightforwardly the anomalies of  CFT$_{UV}$ ($a_{UV},c_{UV}$) and CFT$_{IR}$ ($a_{IR},c_{IR}$) is the operatorial anomaly in
$(T^{(matter)})_\mu^\mu$.  However, the operatorial anomaly can be very easily removed with the aid of a dilaton (alternatively, a conformal compensator). Denoting $\Omega\equiv e^{-\tau}$, we replace every mass scale according to $M_i\rightarrow M_i\Omega$.
 We also add a kinetic term for this dilaton (replacing the metric in~\effact\ by the flat metric) such that the theory becomes
\eqn\theory{S=S_{matter}[\Phi_i,M_i\Omega]+f^2\int d^4x (\del\Omega)^2~.} This theory is now void of operatorial Weyl anomalies, in other words, this theory satisfies the operator equation
\eqn\noopano{T_\mu^\mu=0~.}
Operator equations, by definition, hold at separated points, and this is how~\noopano\ is to be interpreted.

So far we have not said much about the dimensionful scale $f$.
Since it appears as the coefficient of the kinetic term of the dilaton, we see that the physical dilaton fluctuations couple to matter fields by inverse powers of $f$ and thus if we take
\eqn\limit{M_i\ll f~,}
the coupling between the dilaton and the matter sector is arbitrarily weak.

To recapitulate, we have seen that the operator anomaly can be canceled with a dilaton whose fluctuations couple weakly to the matter fields. Thus, setting the dilaton to its VEV ($\Omega=1$) the original matter theory is recovered, and it flows as depicted in Fig.2, perturbed only by the infinitesimal coupling to the dilaton field. Hence, the deep IR theory consists of CFT$_{IR}$ supplemented by the decoupled dilaton field.\foot{This procedure we have invoked for canceling the operatorial anomaly has analogs in many other contexts. In general, every explicit symmetry breaking can be reinterpreted as spontaneous breaking by adding a massless field with an arbitrarily large decay constant (and hence weakly coupled to the matter fields, such that the essential dynamics is intact).  }

Our theory~\theory\ also needs to be defined properly at high energies. We can introduce a cutoff $\Lambda_{UV}$ which satisfies $\Lambda_{UV}\gg M_i$. All momenta are restricted to satisfy $p^2\ll\Lambda^2_{UV}$. We will now see that at high energies (namely $M^2_i\ll p^2\ll\Lambda_{UV}^2$) our system~\theory\ consists of CFT$_{UV}$ plus the additional dilaton, very weakly coupled to CFT$_{UV}$. Indeed, this is true since in the limit~\limit\ the physical dilaton interactions with operators in the conformal theory are various marginal  operators suppressed by $M_i/f$. Consequently, the fact that these operators are not necessarily exactly marginal plays no role to leading order. In other words, there are logarithms which are higher order in $M_i/f$, and since we work to leading order in this expansion, it is consistent to treat the ultraviolet theory as consisting of CFT$_{UV}$ plus a decoupled dilaton.

To summarize, in the limit~\limit\ the introduction of the dilaton modifies the flow in Fig.2 in a trivial way. The UV now effectively consists of CFT$_{UV}$ together with a dilaton, and analogously, the IR consists of CFT$_{IR}$ plus the dilaton. In between the UV ($M^2_i\ll p^2\ll\Lambda_{UV}^2$) and the IR ($p^2\ll M_i^2$) the flow proceeds as in Fig.2, essentially unperturbed by the dilaton.

However, the presence of this innocuous dilaton gives us an important handle on anomalies. Since our theory now satisfies the Ward identity~\noopano, the total $a$- and $c$-trace anomalies of~\theory\ must match between the UV and IR. As in section~3, this does not mean that the anomalies of CFT$_{UV}$ and CFT$_{IR}$ match, rather, that together with the dilaton the total anomaly agrees. Indeed, since along the flow the dilaton is weakly coupled to the matter theory, various effective operators involving the dilaton field are generated upon integrating out the matter fields.

Since we assume~\limit, we are only interested in the leading terms in $1/f$. To leading order in this expansion it is sufficient to integrate out the matter fields while the dilaton sits on external lines. (Internal lines of the dilaton unavoidably suppress the diagrams by further powers of $1/f$.)  Then, the diagrams one needs to compute in order to find the dilaton couplings at low energies  are depicted schematically in Fig.3.

\medskip
\epsfxsize=3.0in \centerline{\epsfbox{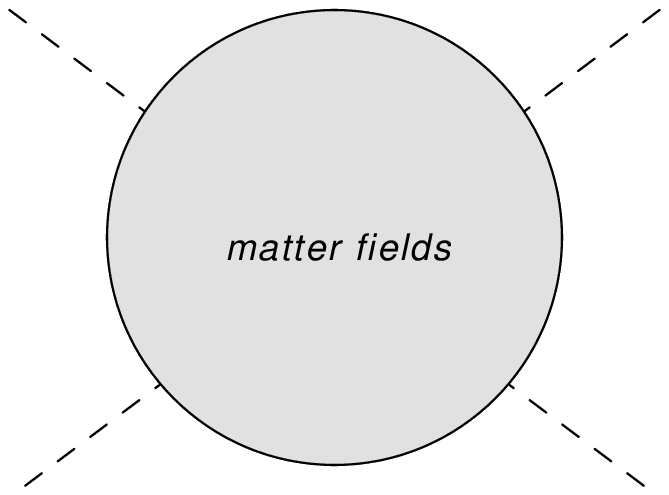}}
\noindent
Fig.3: Terms in the effective action of the dilaton at leading order in $1/f$ are obtained by computing diagrams where the blob consists of matter fields only.  \medskip

The matching of the total anomaly constrains the effective action
in the IR. To make it a little clearer where the different terms
come from we introduce some background metric $g_{\mu\nu}$. Up to
terms with four derivatives acting on the dilaton the most general
allowed effective theory is
\eqn\IRactioni{\eqalign{S_{IR}[g_{\mu\nu}]&={\rm {CFT}}_{IR}[g_{
 \mu\nu}]+{1\over 6}f^2\int d^4x\sqrt{- \hat
g}\hat R+{\kappa\over 36}\int d^4x\sqrt {-\hat g}\hat R^2+\kappa'\int d^4x\sqrt {-\hat g}\hat W_{\mu\nu\rho\sigma}^2\cr&-(a_{UV}-a_{IR})\int d^4x\sqrt{- g}\biggl(
 \tau E_4+ 4\bigl(R^{\mu\nu}-\half g^{\mu\nu}
R\bigr)\del_\mu \tau\del_\nu \tau-4(\del
\tau)^2\square\
 \tau+2(\del \tau)^4\biggr)\cr&+(c_{UV}-c_{IR})\int d^4x \sqrt{-g}  \tau
W_{\mu\nu\rho\sigma}^2~.}} The coefficients of the anomalous
functional are fixed so that the total anomaly of~\IRactioni\
matches the UV anomalies $a_{UV},c_{UV}$. In contrast to
section~3, the dilaton is present both in the UV and in the IR,
hence, its contribution to the anomalies via loop diagrams cancels
from the matching. Note that we have not included a cosmological
constant term, although this is generally generated by the flow.
Indeed, the physical cosmological constant can always be tuned to
zero by including an appropriate bare vacuum energy term
in~\theory. (In addition, the coupling of the dilaton to matter
may affect the normalization of the kinetic term of the dilaton,
but to not clutter the notation we have denoted by $f$ the
physical decay constant already in~\theory).

Since at this stage the physics is identical to what we have
already discussed in great length in section~3, the consequences
are similar too. Namely, the difference between the $a$-anomalies
is isolated by the leading contribution to the $2\rightarrow 2$
$S$-matrix element at low energies. In the forward kinematics, the
scattering amplitude at low energies is given by
\eqn\scatampli{\CA(s)= {2(a_{UV}-a_{IR})\over f^4}s^2+\CO(s^4)~.}

This leads to the sum rule
\eqn\direli{a_{UV}-a_{IR}={f^4\over \pi}\int_{s'>0} ds' {\sigma (s')\over s'^2}~,}
with $\sigma(s')$ the (manifestly positive-definite) cross section for the scattering of two dilatons.
Note that the cross section goes as $1/f^4$ in the limit~\limit, but the prefactor $f^4$ in~\direli\ ensures the result stays finite (as it should) no matter how large $f$ is.
We can also construct an obvious interpolating function which is monotonically decreasing, as in~\decrease\
\eqn\decrease{a(\mu)\equiv a_{UV}-{f^4\over \pi}\int_{s'>\mu}
ds'{\sigma(s')\over s'^2}~.}

This completes the proof of the $a$-theorem. An explicit example that demonstrates the ideas in this section is worked out in appendix B.

\newsec{Discussion and Open Questions}

Recall that in two dimensions there is a ``dispersive''
proof of the $C$-theorem due to~\CappelliYC. Some aspects of our result look analogous to the
corresponding discussion in two dimensions, especially, the role played by the sum rule.
There have been many previous attempts to utilize various sum rules and dispersion relations to shed light on Cardy's conjecture,
but one crucial difference between our approach and previous attempts
is that four-point functions play a pivotal role in our discussion. The scattering of $2\rightarrow 2$ dilatons, contains, after all, data from the correlator of four traces of the stress tensor $\langle T_\mu^\mu T_\mu^\mu T_\mu^\mu T_\mu^\mu\rangle$. Such objects have nice positivity properties, unlike three-point functions (which are commonly used to define the $a$- and $c$-anomalies). It would be instructive to reformulate our proof in the language of correlation functions, never referring to the auxiliary dilaton.\foot{In the case of the chiral anomaly, one
can indeed avoid the spectator fields altogether and analyze the
correlation functions and contact terms in great
detail~\refs{\FrishmanDQ,\NirSV}. }

Let us now allude briefly to some additional questions our analysis raises:
\bigskip

\item{1)} One central ingredient is the ``non-Abelian'' structure of
the Euler anomaly. This leads to the universal $2\rightarrow 2$ scattering
in flat space. It would be interesting to understand better the
algebraic (cohomological) structure of this phenomenon.

\item{2)} We have constructed a monotonic decreasing function that interpolates between $a_{UV}$ and $a_{IR}$. However, we have not addressed the question of whether the RG flow is a gradient flow or not. The four-dilatons coupling~\minimalanomalyi, which is the hero of our story, is obviously related to the four-point function of $T_\mu^\mu$ . Hence, one may speculate that the evolution from the UV to the IR is associated to
a positive-definite quartic differential, rather than the gradient
flow in two dimensions. Being a fundamental property of
four-dimensional RG flows, this is clearly worth addressing. The study of simple examples could be instrumental in attacking this question.

\item{3)} We have not tried to make contact with Cardy's original proposal for $\int_{S^4} \langle
T_\mu^\mu\rangle$ as the quantity that monotonically decreases along the flow. Of
course, in the UV and IR this object coincides with
$a_{UV},a_{IR}$, respectively. But can it be related to our construction at some intermediate scale $\mu$ as well?
Note that our discussion of the $a$-anomaly uses local methods (i.e. an effective action approach), while the integral over the sphere is a global quantity. A better understanding of the relationship between these two approaches is sorely needed.

\item{4)} It would be interesting to understand the effective action of the dilaton on the moduli space of $\CN=4$ Yang-Mills theory (see~\BuchbinderJN\  for a discussion of some aspects of this problem) and to compare with expectations from strong coupling~\TS.

\item{5)} The flat space-time dilaton self-interaction coming from~\minimalanomaly\ may have an interesting manifestation in holography.
One could consider conformal symmetry breaking in AdS spaces and
try to identify the (perhaps geometric) reason for the
universality of the coefficient of this interaction.\foot{One can
verify that in the case of conformal symmetry breaking by a $D3$
brane localized in the radial direction of $AdS_5$, the correct
four-derivative dilaton interaction is captured by the DBI action.
We thank O.~Aharony, J.~Maldacena, and S.~Theisen for pointing this
out to us.}

\item{6)} While the generalization of our results to trace anomalies in 6d~\refs{\BonoraCQ,\DeserYX} seems feasible, the question of what happens when the number of dimensions is odd remains open. A proposal for a measure of degrees of freedom in 3d has been given by~\MyersTJ, and some further evidence for it in supersymmetric theories is discussed in~\JafferisZI. Such a theorem in three dimensions could have applications for condensed matter systems. The conjecture in three dimensions again concerns itself with the partition function over the round three-sphere. We certainly cannot prove this theorem yet, but we would like to offer some intuition for why such a result is not inconceivable.\foot{We thank D.~Jafferis for a crucial discussion of these matters.}  If a quantity is to decrease in every RG flow, it must be constant on conformal manifolds.\foot{This is satisfied by $C$ in two dimensions and also by $a$ in four dimensions. The latter follows trivially from our general results.  Another proof of this is given by repeating verbatim the argument given in the text, replacing $R^3,S^3$ by $R^4,S^4$ (and taking a derivative with respect to $\log R$).} Changing the location on the conformal manifold corresponds to taking a derivative with respect to a coordinate on the conformal manifold, thus one has to calculate a one-point function on $S^3$. But since $S^3$ can be stereographically projected onto $R^3$, one-point functions on $S^3$ of all the primary operators (besides the unit)  are zero. Hence, the partition function on the sphere is constant  on conformal manifolds.

\goodbreak
\bigskip
\bigskip
\centerline{\bf Acknowledgments }

We would like to thank O.~Aharony, M.~Buican, S.~Deser,
T.~Dumitrescu, D.~Z.~Freedman, D.~Jafferis, D.~Kutasov,
J.~Maldacena, R.~Myers, N.~Seiberg, and S.~Theisen  for helpful
conversations. Z.K. is supported by a research grant from Peter
and Patricia Gruber Awards.  Z.K also gratefully acknowledges
support from DOE grant DE-FG02-90ER40542. A.S. acknowledges the
research center supported by the Israel Science Foundation (grant
number 1468/06). Z.K and A.S. would like to thank the organizers
of ``The 6th Regional Meeting in String Theory'' for providing a
great ambience during the beginning of this project.

\appendix{A}{Conventions}
The signature we use is $(+,-,-,-)$. The energy momentum tensor can be extracted from the action by using \eqn\emtensor{T^{\mu\nu}={-2\over \sqrt{-g}}{\delta S\over \delta g_{\mu\nu}}~,\qquad T_{\mu\nu}={2\over \sqrt{-g}}{\delta S\over \delta g^{\mu\nu}}~.} This implies that under Weyl variations, $\delta g_{\mu\nu}=2\sigma g_{\mu\nu}$,
 \eqn\Weyltr{\sqrt{-g}T_\mu^\mu=-{\delta S\over \delta \sigma} ~.}

We also define
\eqn\Riemann{R_{\lambda\mu\nu}^\rho=\del_\mu\Gamma_{\lambda\nu}^\rho-\del_\nu\Gamma_{\mu\lambda}^\rho+\Gamma_{\mu\kappa}^{\rho}\Gamma_{\nu\lambda}^\kappa-\Gamma_{\nu\kappa}^{\rho}\Gamma_{\mu\lambda}^\kappa~,}
which can be contracted to give $R_{\lambda\nu}=R_{\lambda\mu\nu}^\mu$, $R=g^{\lambda\nu}R_{\lambda\nu}$. The Euler tensor is defined as
\eqn\Euler{E_4=R_{\mu\nu\rho\sigma}^2-4R_{\mu\nu}^2+R^2~,} and the Weyl tensor squared satisfies
\eqn\Weylsquared{W_{\mu\nu\rho\sigma}^2=R_{\mu\nu\rho\sigma}^2-2R_{\mu\nu}^2+{1\over 3}R^2~.}

The trace anomaly is then given by
\eqn\trace{T_\mu^\mu= aE_4-cW^2~.}
Real scalars contribute to the anomalies $(a,c)={1\over 90(8\pi)^2}(1,3)$, Weyl fermion: $(a,c)={1\over 90(8\pi)^2}(11/2,9)$, gauge field: $(a,c)={1\over 90(8\pi)^2}(62,36)$.
(The calculation leading to these values is reviewed in~\BirrellIX, where additional references can be found too.)
The variation of the action is then given by \eqn\actvar{\delta_\sigma S=\int d^4x \sqrt{-g}\sigma  \left(cW^2-aE_4\right)~,} with the positive values of $a,c$ quoted above.

\appendix{B}{A Free Massive Field}

As an illustration of the procedure outlined in section~4, we discuss
the flow of a free massive field, coupled to the ``compensator'' (dilaton).

The action is simply
 \eqn\massive{S=\int d^4x\left(\half\del_\mu\Phi
\del^\mu\Phi-\half M^2 \Phi^2\right )~.}
We introduce a dimensionless compensator field $\Omega$, with expectation value $\Omega=1$ such that the action becomes
\eqn\comp{S=\int d^4x
\left(f^2\del_\mu\Omega\del^\mu\Omega
+\half\del_\mu\Phi\del^\mu\Phi-\half M^2\Omega^2\Phi^2\right)~.}
The action~\comp, which is now interactive, should be understood in the
presence of an ultraviolet cutoff $\Lambda$ . We will
restrict always all the momenta $p$ to the range
 $p^2\ll\Lambda^2$. We would not like the compensator to modify the flow of the free massive field, so we take
 \eqn\lrange{f\gg M~.} In this limit the physical coupling between $\Omega$ and $\Phi$ is arbitrarily weak.

With these assumptions, at large momenta the theory behaves  as a conformal theory since the ultraviolet logarithms (due to the quartic interaction)
are suppressed by positive powers of $M/f$. We will always compute quantities only to leading order in $M/f$ and so the theory can be treated as if it is exactly conformally invariant at energies
$M\ll E\ll (\Lambda,f)$. (We remark that such an embedding of the free massive field in a conformal theory is explicitly
realized on the Coulomb branch of the $\CN=4$ supersymmetric gauge
theory at weak coupling where the range $p^2\ll\Lambda^2$ is extended to
infinity since the ultraviolet logarithms cancel due to
supersymmetry.)

We proceed now to study the broken phase of the theory,
i.e. when $\Omega $ gets a vacuum expectation value
\eqn\vev{<\Omega>=1~.}
Expanding the field $\Omega$ around the VEV, we
define the fluctuation $\varphi$ through \eqn\fluct{\Omega=e^{-\tau}=1-\varphi~.}
The action becomes now:
 \eqn\brokact{S=\int d^4x \left(f^2\del_\mu\varphi\del^\mu\varphi +\half \del_\mu\Phi\del^\mu\Phi-\half
M^2\Phi^2-M^2 J\Phi^2\right )~,}
where \eqn\defJ{J=\left(\half\varphi^2-\varphi \right)~.}

We are interested in the effective action of the $\varphi$ field after
integrating out the massive $\Phi$ . Since the action is quadratic in $\Phi$ this is exhausted by the  one-loop
diagrams of Fig.4. Our limit~\lrange\ guarantees that diagrams where the $\varphi$  circulates
in the loops are suppressed and can thus be neglected.

\medskip
\epsfxsize=3.5in \centerline{\epsfbox{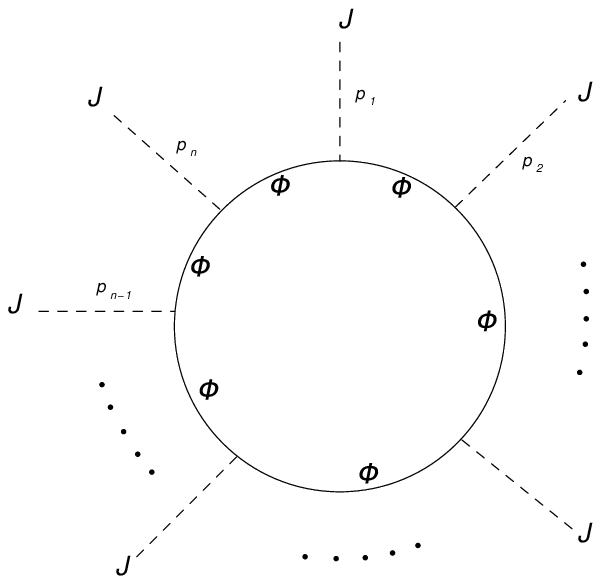}}
\noindent
Fig.4: We integrate out the massive scalar $\Phi$ and obtain an effective action for $J$, which can then be translated to $\varphi$ via~\defJ.  \medskip

The infrared action of the $\varphi$ field is given by an expansion in powers of the momenta
$p^2$.
In particular, we are interested in the $p^4$ scale independent
terms which should be related to the anomaly.

We illustrate the method of our calculation for the simplest  $J^2$
term . The prototypical Feynman integral with two external momenta $p_1,p_2$ is
given by:
\eqn\feyn{M^4
\delta^{(4)}\left(p_1+p_2\right)\int {d^4q \over
\left(q^2-M^2\right)\left[\left(q-p_1\right)^2-M^2\right]}~.}
One combines the propagators using the Feynman
trick and expands to order $p^4$. The remaining ultraviolet convergent integral is
\eqn\expr{3
M^4\left(p_1^2\right)^2 \delta\left(p_1+p_2\right)
\int_0^1d\alpha \alpha^2\left(1-\alpha\right)^2 \int{d^4 q \over
\left(q^2-M^2\right)^4} ~.}
This evaluates to  \eqn\fin{i{\pi^2 \over
60} \left(p_1^2\right)^2~.}

Taking into account all the symmetry factors, one finds that in order to reproduce this four-derivative $J^2$ term via a contribution to the effective action we must include
\eqn\effeac{{\pi^2 \over 60\left(2\pi\right)^4}J {\square }^2 J~.}
Similarly, we calculate the Feynman diagrams associated to three and four external $J$ fields. We expand these diagrams to four derivatives. Substituting~\defJ\ we get the four-derivative effective action for~$\varphi$ (neglecting terms with more than four $\varphi$s)  \eqn\final{  {1 \over
2880\pi^2}  \left [  \left( \left
(   \nabla\varphi  \right)^2 \right)^2+9\varphi^2\left(\square \varphi\right)^2
+6\varphi\left(\nabla\varphi\right)^2 \square \  \varphi+2
\left(\nabla\varphi\right)^2\square \ \varphi
+6\varphi\left(\square \ \varphi\right)^2 +3
\left(\square \ \varphi\right)^2\right]~.}

The expression~\final\ fits the terms we expect from~\IRactioni\
(recall~\canonical). There is a contribution from the invariant
piece proportional to $\kappa$ in~\IRactioni\ and also the terms
associated to the anomaly functional are present. Indeed, the
theory~\massive\ evolves from a CFT$_{UV}$ containing a free
massless field to an empty theory in the IR. So we expect to
obtain~\IRactioni\ with \eqn\ano{a_{UV}-a_{IR}={1 \over 5760
\pi^2}~,} which is exactly the right value contained in~\final.

As in the general construction of section~4, the change in the $a$-anomaly of a free massive field~\ano\ can be represented by the dispersion relation~\direli. The integral over the branch cut runs from the threshold $4m^2$ to infinity and one can construct the interpolating function~\decrease\ explicitly.

\listrefs
\end